\documentclass[a4paper,fleqn,usenatbib]{mnras}

\usepackage{mathptmx}
\usepackage[T1]{fontenc}
\usepackage{ae,aecompl}

\usepackage{graphicx}	% Including figure files
\usepackage{amsmath}	% Advanced maths commands
\usepackage{amssymb}	% Extra maths symbols
\graphicspath{{Figures/}}   % Path to figures
\usepackage[final]{changes}      %changes

\title[Impact of baryons and the selection effect]{Super-cluster simulations: impact of baryons on the matter power spectrum and weak lensing forecasts for Super-CLASS}

\author[Aaron Peters et al.]{
Aaron Peters,$^{1}$\thanks{Contact email: aaron.peters@postgrad.manchester.ac.uk}
Michael L. Brown,$^{1}$
Scott T. Kay$^{1}$
and David J. Barnes$^{1}$
\\
$^{1}$Jodrell Bank Centre for Astrophysics, School of Physics and Astronomy, The University of
Manchester, Manchester M13 9PL}

\date{Accepted XXX. Received YYY; in original form ZZZ}

\pubyear{2016}

\begin{document}
\label{firstpage}
\pagerange{\pageref{firstpage}--\pageref{lastpage}}
\maketitle

\begin{abstract}
We use a combination of full hydrodynamic and dark matter only simulations to investigate the effect that baryonic physics and selecting super-cluster regions have on the matter power spectrum, by re-simulating a sample of super-cluster sub-volumes. On large scales we find that the matter power spectrum measured from our super-cluster sample has at least twice as much power as that measured from our random sample. Our investigation of the effect of baryonic physics on the matter power spectrum is found to be in agreement with previous studies and is weaker than the selection effect over the majority of scales. In addition, we investigate the effect of targeting a cosmologically non-representative, super-cluster region of the sky on the weak lensing shear power spectrum. We do this by generating shear and convergence maps using a line of sight integration technique, which intercepts our random and super-cluster sub-volumes. We find the convergence power spectrum measured from our super-cluster sample has a larger amplitude than that measured from the random sample at all scales. We frame our results within the context of the Super-CLuster Assisted Shear Survey (Super-CLASS), which aims to measure the cosmic shear signal in the radio band by targeting a region of the sky that contains five Abell clusters. Assuming the Super-CLASS survey will have a source density of 1.5 galaxies/arcmin$^2$, we forecast a detection significance of $2.7^{+1.5}_{-1.2}$, which indicates that in the absence of systematics the Super-CLASS project could make a cosmic shear detection with radio data alone. 
\end{abstract}


%%%%%%%%%%%%%%%%%%%%%%%%%%%%%%%%%%%%%%%%%%%%%%%%%%

%%%%%%%%%%%%%%%%% BODY OF PAPER %%%%%%%%%%%%%%%%%%

\section{Introduction}
Light from background source galaxies propagating through 
the Universe gets deflected by inhomogeneities in the matter distribution. 
This phenomenon, termed weak gravitational lensing, 
causes coherent distortions, typically 
percent level, 
to both the size (convergence) 
and shape (shear) of source galaxies 
\citep[for reviews see,][]{Bartelmann:1999yn,Schneider:2005ka,Hoekstra:2008db}. 
Weak gravitational lensing, when  
applied to studies of the large scale structure of the Universe, is known as 
cosmic shear and can be directly related to the 
matter power spectrum \citep{1992ApJ...388..272K}. 

Cosmic shear measurements have been successfully used to constrain 
cosmological parameters in various surveys 
\citep[some recent results include e.g.][]{refId0,2013ApJ...765...74J,2013MNRAS.431.1547B,2013MNRAS.430.2200K,Heymans:2013fya,Fu:2014loa,PhysRevD.94.022001,Hildebrandt:2016iqg}, 
and are believed to hold the most promise 
for probing dark energy. 
In addition, some of the most 
accurate constraints on the matter power spectrum come from weak lensing, 
particularly in the non-linear regime. 
The next stage of optical weak lensing surveys, e.g. 
the ground based Large Synoptic Survey 
Telescope\footnote{http://www.lsst.org/lsst/} 
(LSST) or 
the space mission 
Euclid\footnote{http://sci.esa.int/euclid/}, 
are expected to 
observe two orders of magnitude more source galaxies 
than completed weak lensing surveys. 
As a result, this decrease in 
statistical uncertainty brings about the 
difficult task of reducing the systematic 
uncertainties by an order of magnitude. 
Due to the statistical errors currently present in lensing, analytic models based on 
simulations which assume that baryons trace 
dark matter particles perfectly, i.e., 
dark matter only simulations, have been sufficient. 

However, to take full advantage of future surveys, 
the accuracy with which we model the matter power spectrum 
will have to be improved to the few percent level. 
As a result we must include baryonic processes, such as 
star formation, radiative cooling, supernovae feedback 
and active galactic nuclei (AGN) feedback. 
Cooling gas has been found to cluster on smaller scales than 
dark matter, which means the matter power spectrum 
is under-predicted in dark matter only simulations on scales 
greater than $k \sim 6 \: h/$Mpc 
\citep[e.g.,][]{2006ApJ...640L.119J,2008ApJ...672...19R,2010MNRAS.405..525G,2011MNRAS.412..911C,2011MNRAS.415.3649V}. 
Additionally, \citet{2011MNRAS.415.3649V} found 
that matter power spectra measured in 
hydrodynamic
simulations that model AGN feedback 
were suppressed on intermediate scales, when compared 
to dark matter only simulations.  
In a follow up paper \citep{2011MNRAS.417.2020S}, 
the effect of baryons on the matter power spectrum, 
when used as part of a cosmic shear study, 
was found to significantly bias cosmological parameter estimation. 

%radio weak lensing
In addition, systematic effects can be reduced 
by performing multi wavelength weak lensing investigations \citep{Demetroullas:2015gsa}. 
Particularly, performing 
weak lensing studies in the radio band 
provides a unique opportunity 
to reduce, e.g., systematics due to 
the point spread function and intrinsic 
galaxy 
alignments \citep{2011MNRAS.410.2057B}. 
The smaller fields of view, lower effective source densities 
and limited resolution presently available mean 
the optical waveband is favoured for lensing surveys. 
However, with the upcoming arrival of wide area 
radio surveys, such as the Square Kilometre 
Array (SKA\footnote{https://www.skatelescope.org/}), 
this will change 
\citep{Brown:2015ucq,Harrison:2016stv,Bonaldi:2016lbd}. 
The Super-CLuster Assisted Shear 
Survey's (Super-CLASS\footnote{http://www.e-merlin.ac.uk/legacy/projects/superclass.html})
principal science goal is to measure the 
cosmic shear signal, while further developing techniques for 
analysing radio weak lensing data. 

%Super-cluster fields
In order to maximise 
the cosmic shear signal 
measured, Super-CLASS targets a region of the 
sky known to contain a super-cluster made up of 
five Abell clusters. 
Observations of super-cluster fields from the 
COMBO-17 \citep{2002ApJ...568..141G} and 
STAGES HST surveys \citep{2008MNRAS.385.1431H}, 
used weak lensing analysis to probe the dark 
matter distribution of the Abell 901/902 super-cluster. 
In addition, \citet{2003MNRAS.341..100B} found the shear power spectrum 
measured from the Abell 901/902 super-cluster field had an 
amplitude significantly higher than that measured 
in random fields. 

We know small super-cluster fields are not cosmologically 
representative regions of the sky, which implies weak lensing measurements are 
going to be affected when selecting a super-cluster field as opposed to a random field. 
In this paper we intend to quantify 
this effect, hereafter referred to as the `selection effect', 
as well as the effect of baryonic physics on the matter power spectrum. 
We do this by selecting 
a sample of 61 super-cluster sub-volumes 
that meet criteria specifically chosen 
to mimic the five Abell clusters targeted 
by the Super-CLASS survey, along with a sample of 60 random sub-volumes, 
from a large volume dark matter only simulation. 
The sub-volumes in these two samples are re-simulated 
using the zoomed re-simulation technique at a higher resolution 
and with full gas physics. With these re-simulations we 
are able to compare the difference between measuring 
the matter power spectrum in random and super-cluster sub-volumes. 
In addition, we investigate how the matter power spectrum is 
affected by baryonic physics, and which of the two effects is dominant. 
We also generate shear and convergence maps using a line 
of sight integration technique, which intercept the random 
and supercluster sub-volumes at $ z = 0.24 $, approximately matching the Super-CLASS
super-cluster redshift. These are then used 
to study the difference between the 
shear power spectrum measured by 
a weak lensing survey that targets a random patch of the sky 
as opposed to one containing a super-cluster. 

%outline of paper
The outline of this paper is as follows. 
We discuss the simulations used for this work 
and the identification method employed to select 
a sample of super-cluster sub-volumes, 
in Section \ref{sec:simulations}. 
In Section \ref{sec:matterpowerspectrum} we examine 
how the selection effect will affect the matter 
power spectrum (Section \ref{sec:environment}), 
we investigate how baryonic process 
affect the matter power spectrum (Section \ref{sec:baryons}), 
and we also compare the effect of baryonic physics on the matter 
power spectrum to that of selection (Section \ref{sec:environmentbaryons}). 
In Section \ref{sec:weaklensing} 
we detail the line of sight integration 
technique used to generate convergence 
and shear maps (Section \ref{sec:lightcones}), 
%and 
the effect of selection on the shear and convergence 
power spectra (Section \ref{sec:environmentweaklensing}) 
and lastly we forecast the 
constraining power of the 
Super-CLASS survey (Section \ref{sec:forecasts}). 
Finally, we conclude by summarising 
our results in Section \ref{sec:conclusion}. 

\section{Simulations}
\label{sec:simulations}
In the following section we describe the large 
dark matter only simulation we used, hereafter referred to as the ``parent'' simulation. 
Additionally, we detail the selection criteria used to identify the super-cluster sample 
and the baryonic physic implemented when we 
re-simulated the sample sub-volumes with the zoom technique. 

\subsection{Parent Simulation}
\begin{table*}
\caption{Key observational properties of the five Abell 
clusters in the Super-CLASS field. Luminosities are taken from the BAX 
database \citep{2004A&A...424.1097S}, while the positions on the 
sky and redshifts are taken 
from \citet{1993A&A...278..379B}. Masses are estimated from 
the $L_{\mathrm{x}}-M_{500}$ relation (see text for details). }
\label{tab:Abell}
\begin{tabular}{lccccc}
\hline
Cluster Name & RA 1950 & DEC 1950 & $z$ & $L_{\mathrm{x}}$ $[0.1-2.4]$keV & $M_{500}$ \\
\hline
Abell 968 & 10 17 44.1 & 68 36 34  & 0.195 & $0.401 \times 10^{44}$ erg/s & $(1.2 \pm 0.3) \times 10^{14} \: {\mathrm{M}}_\odot$ \\
Abell 981 & 10 20 36.0 & 68 20 06 & 0.201 & $1.670 \times 10^{44}$ erg/s & $(2.7 \pm 0.7) \times 10^{14} \: {\mathrm{M}}_\odot$ \\
Abell 998 & 10 22 47.8 & 68 11 13 & 0.203 & $0.411 \times 10^{44}$ erg/s & $(1.2 \pm 0.3) \times 10^{14} \: {\mathrm{M}}_\odot$ \\
Abell 1005 & 10 23 40.0 & 68 27 18 & 0.200 & $0.268 \times 10^{44}$ erg/s & $ (1.0 \pm 0.2) \times 10^{14} \: {\mathrm{M}}_\odot$ \\
Abell 1006 & 10 24 10.7 & 67 17 44 & 0.204 & $1.320 \times 10^{44}$ erg/s & $ (2.4 \pm 0.6) \times 10^{14} \: {\mathrm{M}}_\odot$ \\
\hline
\end{tabular}
\end{table*}
A simulation with a very large cosmologically representative volume, 
of several Gpc$^3$, is necessary to simulate the large population 
of massive clusters that we require to find a sample of super-clusters. 
To investigate the effect of baryons in super-cluster regions 
we require high resolution hydrodynamic simulations with 
volumes that are sufficiently large to produce the large scale 
power needed for a sample of super-clusters to be identified. 
Computational resources restrict our ability to 
simulate hydrodynamical volumes of this magnitude to a sufficiently high resolution 
given the gas physics we wish to implement. 
However, we can apply a workaround called 
the zoomed re-simulation 
technique \citep{1993ApJ...412..455K,1997MNRAS.286..865T}, where 
a sub-volume of a larger dark matter only simulation is 
re-simulated at higher resolution. 
We will use this technique to re-simulate 
sub-volumes of the parent simulation for our matter power spectrum 
investigation. 
In addition, the large volume of the parent simulation 
was needed for the weak lensing analysis, 
to generate the shear and convergence maps along light-cones through 
the simulation. 

The parent simulation is a large periodic dark matter only simulation 
with a cubic volume of (3.2 Gpc)$^3$, which we used to select 
our sample of super-clusters. 
It follows the evolution of 2520$^3$ dark matter particles, 
each of mass $5.43 \times 10^{10} \: {\mathrm{M}}_\odot /h$, from 
$ z = 127 $ to $ z = 0 $. 
The comoving gravitational softening length 
of the parent simulation is 40 kpc/$h$. 
The simulation was run using 
the $\textsc{gadget}$-3 code \citep{2008MNRAS.391.1685S}, 
an updated and more efficient version of the publicly 
available $\textsc{gadget}$-2 code \citep{Springel:2005mi}. 

%initial conditions
The initial conditions were created by first 
generating a glass like particle distribution \citep{White:1994bn}, 
then each particle's displacement 
and velocity were calculated 
according to second order perturbation theory using 
the method laid out in 
\citet{2010MNRAS.403.1859J}, using the public 
Gaussian white noise field 
$\textsc{panphasia}$ \citep{2013MNRAS.434.2094J,2013arXiv1306.5771J}. 
Its cosmological parameters 
$[\Omega_\mathrm{m}, \Omega_{\Lambda}, \Omega_\mathrm{b} , h, \sigma_8, n_s, Y ] =\,\,$[0.307, 0.693, 0.04825, 0.6777, 0.8288, 0.9611,0.248], 
were chosen to be consistent with Planck 
year 1 results \citep{2014A&A...571A..16P}. 

\subsection{Supercluster Identification}
\label{subsec:superclusteridentification}
Here we detail the method used to identify a sample 
of super-cluster sub-volumes within the parent simulation. 
As our aim is to identify super-cluster sub-volumes 
with properties that best resemble 
the five Abell clusters in the Super-CLASS field, we 
begin with a brief description of this. 

%Super-CLuster Assisted Shear Survey 
The Super-CLASS project is a deep field, radio weak lensing survey. 
The project's primary science goal is to detect a cosmic shear signal 
by targeting a region of sky containing a super-cluster. 
Specifically, the 
one square degree, Super-CLASS field contains five 
known, $z\sim0.2$ 
Abell clusters (A968, A981, A998, A1005, A1006). 
We detail in Table \ref{tab:Abell} some of their observational properties. 
The positions on the sky and redshifts were taken from \citet{1993A&A...278..379B}, 
while the X-ray luminosities were taken from the BAX database \citep{2004A&A...424.1097S}. 
\citet{Pratt:2008bf} investigated X-ray luminosity scaling relations for a sample 
of 33 local ($z < 0.2$) galaxy clusters from REXCESS. 
We determined the $M_{500}$\footnote{We define $M_{500}$ 
as the mass enclosed within a sphere of radius 
$R_{500}$, in which the mean density is five hundred times that of 
the critical density of the Universe.} 
masses listed in the final column of Table \ref{tab:Abell} 
using the following relation from the aforementioned paper 
\begin{equation}\label{eq:luminosity}
M_{500} = \left (\dfrac{L_x}{L_0} h(z)^{-7/3}  \right )^{1/\alpha} M_0,
\end{equation}
where $h^2(z) = \Omega_{\mathrm{m}}(1+z)^3 + \Omega_{\Lambda}$. 
The values of the constants for luminosities $L_x$ 
measured in the $[0.1 - 2.4]$ keV band are as follows, 
$L_0 = 0.78 \pm 0.07 \times 10^{44}$ erg/s, 
$\alpha = 1.83 \pm 0.14$ and $M_0 = 2 \times 10^{14} \: {\mathrm{M}}_\odot$. 

To select similar regions in the simulation, 
we first identified 
clusters 
at the snapshot that 
corresponds to $z=0.24$, 
the closest available redshift. 
Halo structures were identified using the 
$\textit{Friends-of-Friends}$ \citep[$\textsc{fof}$,][]{1985ApJ...292..371D} algorithm  
on particles, choosing a linking length of $b=0.2$ times 
the mean inter-particle separation. 
To separate the $\textsc{fof}$ halo structures into self 
bound substructures the $\textsc{subfind}$ 
algorithm \citep{2001MNRAS.328..726S,2009MNRAS.399..497D} was applied. 
Finally, the $\textit{spherical overdensity}$ 
algorithm \citep[$\textsc{so}$,][]{1994MNRAS.271..676L} was 
applied to each of the $\textsc{subfind}$ haloes, to 
calculate the spherical overdensity radii and masses. 

The criteria listed below 
were chosen to select a sample of 
sub-volumes in the parent simulation, 
that contain super-clusters with properties 
that best resemble the five Abell clusters 
contained in the Super-CLASS field. 
To identify these regions we ran the $\textsc{fof}$ algorithm 
on haloes, rather than particles. 

\begin{enumerate}
\item 
The $\textsc{fof}$ algorithm was only used on haloes 
that fell in the cluster mass 
range, $(0.5\le M_{500} \le 5) \times 10^{14} \: {\mathrm{M}}_\odot /h$. 
Our mass range was based on the estimated masses of the Abell clusters, 
shown in Table \ref{tab:Abell}. 

\item
The $\textsc{fof}$ linking length $l$, was chosen to 
approximately equal 
the mean projected comoving distance separating the five Abell clusters. 
Rounding to the nearest integer, 
the linking length was determined to be $l = 8$ Mpc/$h$. 

\item
After using the $\textsc{fof}$ algorithm as stated 
we selected the super-clusters 
that contained exactly 5 clusters. 
We identified 61 super-clusters, each containing 
five cluster members that fall in the mass range 
$(0.5\le M_{500} \le 5) \times 10^{14} \: {\mathrm{M}}_\odot /h$ 
and all of which are 
within 8 Mpc/$h$ of at least one other member. 

\end{enumerate}

\begin{figure*}
 \includegraphics[width=\textwidth]{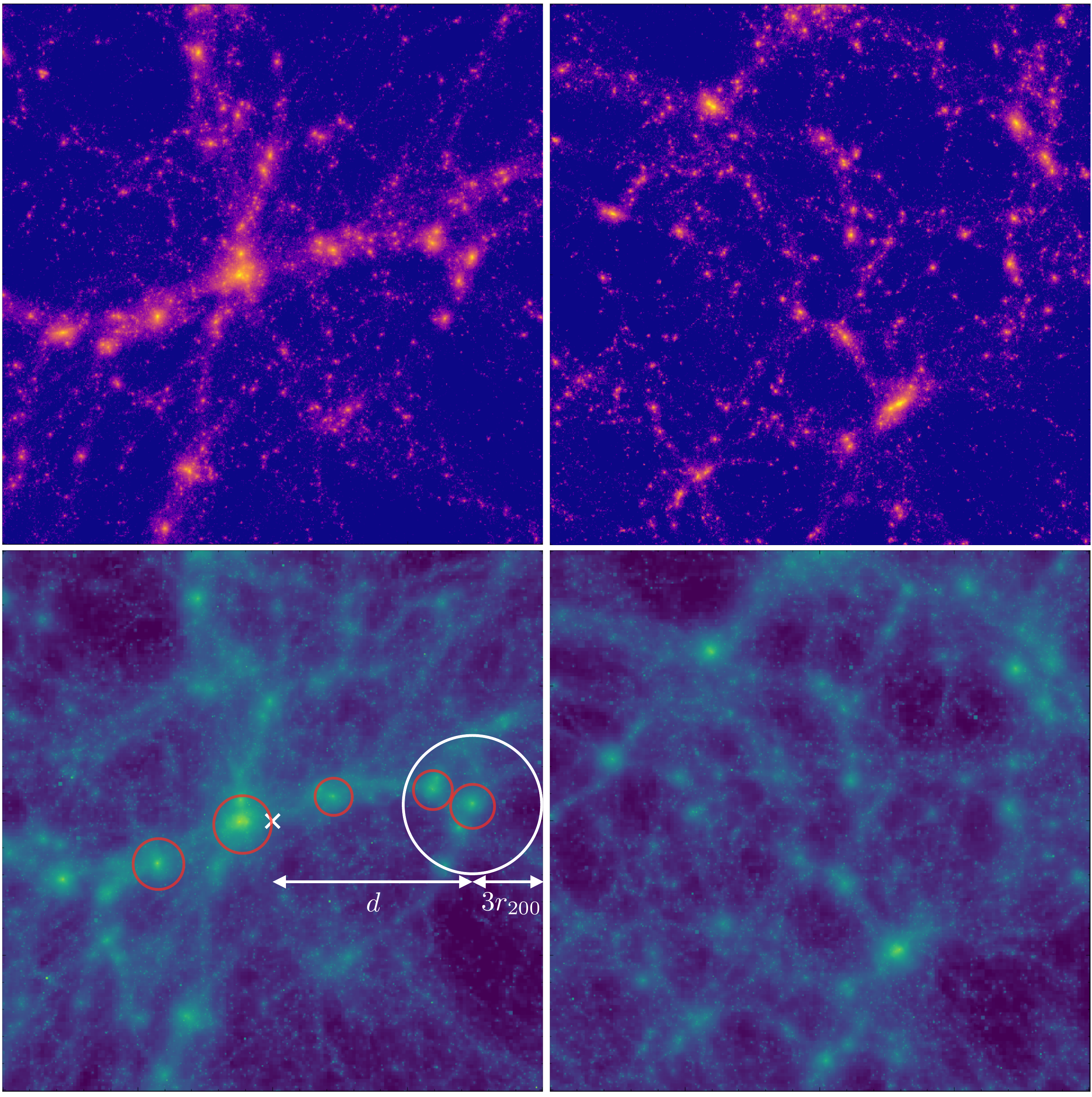}
 \caption{Example of the two dimensional surface mass projection 
 for one super-cluster sub-volume 
 in the left two panels and one random sub-volume in the right two panels. 
 The surface mass projections of the dark matter and gas are shown 
 in the top and bottom panels respectively. 
 The cross in the centre of the bottom left panel is the super-cluster's centre of mass, 
 the 5 red circles with radii of $r_{200}$ enclose the 5 clutser's centre of mass and 
 the white circle with a radius of $3r_{200}$ is drawn about the 
 centre of mass of a particular member cluster. The distance projected along an axis from 
 the centre of mass of the circled cluster to the 
 centre of mass of the super-cluster, $d$, is the longest of any cluster in our super-cluster 
 sample (see text for details).}
 \label{fig:random_supercluster}
\end{figure*}
Finally, the 
minimum side length of a cube 
that could be drawn about any super-cluster's centre of mass and 
guarantee the enclosure of all five respective clusters 
was determined as follows. 
The bottom left panel of Fig. \ref{fig:random_supercluster} 
shows a two dimensional mass projection of the gas in 
the super-cluster 
which has the largest distance $d$ between 
its centre of mass and that of its furthest 
cluster, projected along one of the $x,y,z$ axes. 
Assuming a cluster is completely enclosed by a sphere of 
radius $3r_{200}$ about its centre of mass, a cube drawn 
about the super-cluster's centre of mass with 
a side length of $2(d + 3r_{200})$ will enclose all member 
clusters completely. 
This was found to be 40.16 Mpc/$h$ for the super-cluster shown 
in Fig. \ref{fig:random_supercluster} and, being the most extreme case, 
was used for the rest of the sample to guarantee 
their respective 
enclosure. 

For comparison purposes and to quantify the enhancement 
of the power spectrum in super-cluster regions, we also randomly 
selected 60 additional cubic sub-volumes from the parent simulation. 
These sub-volumes 
were used as a control sample and like the super-cluster sub-volumes 
also have a side length of 40.16 Mpc/$h$. 
Hereafter, we will refer to them as the random sample. 
An example of one of these sub-volumes' 
is also shown in the right panels of Fig. \ref{fig:random_supercluster}. 

\subsection{Zoomed Re-simulations}
\label{sec:zoomresimulations}

%general section intro
In this section we describe the zoomed re-simulation 
technique that we used to re-simulate our sample 
of random and super-cluster sub-volumes 
of the parent simulation. 
These re-simulations are part of the Virgo consortium's 
MAssive Clusters and Intercluster Structures (MACSIS) 
project \citep{2016arXiv160704569B,Henson:2016eip}, 
which extends the 
BAryons and HAloes of MAssive Systems (BAHAMAS) \citep{McCarthy:2016mry} 
simulation to more massive clusters. 

While the aim of this study is different, 
we used the same method employed in the MACSIS project 
to re-simulate the sub-volumes of interest. 
For every member of the random and super-cluster 
samples we used the BAHAMAS code to perform 
both dark matter only and hydrodynamical 
re-simulations, 
using the same cosmological parameters. 
The resolution of the initial conditions were reduced 
throughout the parent simulation except for the sub-volume 
of interest, where they were enhanced. This means 
on re-simulation, the large scale power of the parent 
simulation was preserved and the region of interest 
was simulated at a sufficiently high resolution. 

%describe the gadget owls model
To re-simulate the sub-volumes 
we used 
a heavily modified version of 
the Tree-PM, 
smooth particle hydrodynamical code GADGET-3, 
that was developed to include subgrid physics 
in the suite of simulations used in the OWLS project \citep{2010MNRAS.402.1536S}. 
Some of the subgrid prescriptions included are 
star formation \citep{2008MNRAS.383.1210S}, 
radiative cooling \citep{2009MNRAS.393...99W}, 
feedback from supernovae \citep{2008MNRAS.387.1431D}
and AGN \citep{2009MNRAS.398...53B}. 
\citet{McCarthy:2016mry} showed that by 
calibrating the subgrid model of feedback 
to a small number of observables, 
the BAHAMAS simulation 
was able to reproduce the observed stellar and gas content of massive 
systems. 

%describe the particle masses etc
We performed dark matter only (DMO) re-simulations 
with a particle mass of $5.2 \times 10^{9} \: {\mathrm{M}}_\odot /h$ 
and hydrodynamical (HYDRO) 
re-simulations with a dark matter particle mass of 
$4.4 \times 10^{9} \: {\mathrm{M}}_\odot /h$  and an 
initial gas particle mass of $8.0 \times 10^{8} \: {\mathrm{M}}_\odot /h$, 
for every member 
of the random and super-cluster samples. 
%gravitational softening length
Below $ z = 3 $ 
the gravitational softening length of the high resolution particles 
were set to 4 kpc$/h$ in physical co-ordinates, whilst 
for $ z > 3 $ they were set to 16 kpc$/h$ in co-moving co-ordinates. 

\begin{figure}
 \includegraphics[width=\columnwidth]{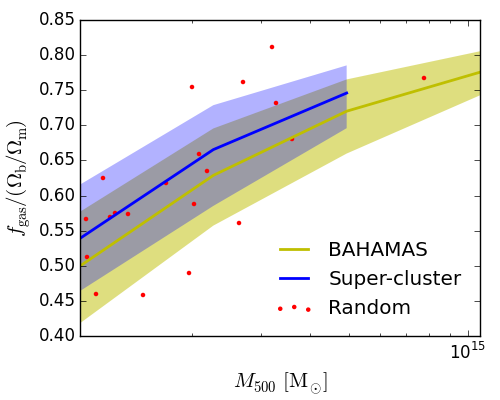}
 \caption{Comparison of the median $f_{\mathrm{gas}}-M_{500}$ relation 
 for clusters from the BAHAMAS sample to 
 the clusters contained in our 
 super-cluster sub-volumes, shown in 
 yellow and blue respectively. 
 The shaded area surrounding the 
 yellow and blue curves represent the region between their 
 $16^{\mathrm{th}}$ and $84^{\mathrm{th}}$ percentiles. 
 We were unable to bin the limited number of massive clusters 
 in our random sample, 
 they are instead shown as red dots.
 In addition the high mass bin of the super-cluster sample has been 
 removed, as it only consisted of 3 clusters. }
 \label{fig:fbarm500}
\end{figure}

In Fig. \ref{fig:fbarm500} we show how the hot gas fraction $f_{\mathrm{gas}}$,  
measured within $R_{500}$, 
depends on their $M_{500}$. 
We calculate $f_{\mathrm{gas}}$ by dividing the total mass of the 
gas with temperature above $10^{5.2}$ K, which we would 
expect to be emitting X-rays, by the total mass of all matter within a radius 
of $R_{500}$ from the centre of the cluster. 
The yellow curve is the 
median $f_{\mathrm{gas}}$ 
for clusters in the BAHAMAS sample, while the blue curve is that taken 
from our super-cluster sample. 
The shaded area surrounding the 
yellow and blue curves represent the region between their 
$16^{\mathrm{th}}$ and $84^{\mathrm{th}}$ percentiles. 
We were unable to bin the massive clusters from our random sample 
due to limited numbers, they are instead shown as red dots. 
Similarly, the high mass bin for the super-cluster sample was removed 
from the plot as it only consisted of 3 clusters. 
The slight discrepancy, between the median curves of the 
BAHAMAS and super-cluster samples, could be due to 
the fact that BAHAMAS uses a 
slightly different cosmology. Multiplying $f_{\mathrm{gas}}$ by the 
universal gas fraction is not necessarily sufficient to correct for this effect. 
However, the more likely cause is the fact that the majority of our clusters are 
situated in overdense 
super-cluster regions. 
These overdense regions 
would most likely make it harder for the gas 
heated by stars and AGN to escape, 
which would result in increased gas fractions in our clusters 
(this is consistent with our findings 
below, in Fig. \ref{fig:pkdmo_overdensity2}).

\section{3D Matter Power Spectrum}\label{sec:matterpowerspectrum}
In this section we will present the matter power spectrum 
estimates measured from our simulations. First, we will 
investigate the selection effect (introduced by specifically 
targeting super-cluster regions) by 
comparing the matter power spectrum measured from our random 
sub-volumes to those of our super-cluster sub-volumes. Then 
we study the effect of baryonic physics on the matter 
power spectrum, before examining which of these 
effects, selection or baryons, 
is more important. 

In order to describe fluctuations in the matter density $\rho(\mathbf{x},t)$, 
at comoving position $\mathbf{x}$, and time $t$,
it is convenient to use the 
overdensity 
\begin{equation}\label{overdensity}
\delta(\mathbf{x},t) = \frac{\rho(\mathbf{x},t)}{\bar{\rho}(t)} - 1,
\end{equation}
where $\bar{\rho}(t)$ is the 
background density of the Universe. 
The overdensity has the range, $-1 \leq  \delta< \infty$, 
with $-1 \leq  \delta< 0$ corresponding to underdense regions 
and $0 < \delta$ corresponding to overdense regions. 
Expanding the overdensity field $\delta(\mathbf{x})$ 
in terms of Fourier components, in a large box of comoving 
volume $V$ we can represent the overdensity field such that 
\begin{equation}\label{overdensityfourier}
\delta(\mathbf{x}) = \dfrac{V}{(2 \mathrm{\pi})^3} \int_V  \delta_{\mathbf{k}}\: \mathrm{exp}(i\mathbf{k}.\mathbf{x}) \: \mathrm{d}^3 k,
\end{equation}
where $\mathbf{k}$ is the comoving wavenumber. 
The Fourier components can be expressed in terms of an 
amplitude $\left | \delta_{\mathbf{k}} \right |$,
and phase $\phi_{\mathbf{k}}$ such that 
$\delta_{\mathbf{k}} = \left | \delta_{\mathbf{k}} \right | \mathrm{exp}(i\phi_{\mathbf{k}})$. 

The overdensity field $\delta (\mathbf{x})$ can be approximated as 
Gaussian 
whilst in the linear regime, when the phases and amplitudes of the 
Fourier components are independent. 
The full statistical properties of a homogenous, isotropic Gaussian 
random field are described by the power spectrum $P(k)$, 
defined to be
$\left \langle \left |\delta_{\mathbf{k}}  \right |^{2} \right \rangle$, where we take 
the average over all possible spatial orientations. Note that the power spectrum 
depends only on the modulus of the wavenumber $\mathbf{k}$ and not its orientation. 

%powmes
Throughout this paper we used the 
publicly available 
code $\textsc{powmes}$\footnote{http://www.projet-horizon.fr/article345.html} 
\citep{Colombi:2008dw} to measure the 
matter power spectrum from our 
simulations. 
$\textsc{powmes}$ estimates the particle distribution's 
Fourier modes, using a Taylor expansion 
on the cosine and sine transforms. 
The accuracy of the Fourier transform increases 
by using higher order approximations of the 
Taylor expansion, taking 
into account small displacements 
within grid cells. For a more detailed description 
of the code see \citet{Colombi:2008dw}. 

Unless stated otherwise, all matter power 
spectra estimates in this paper 
were measured from the snapshot 
corresponding to $z=0.24$, to mimic the properties of 
the Abell clusters in the Super-CLASS field. 
In addition, we found that the shot noise 
due to the discreteness of particles in the DMO and HYDRO sub-volumes 
does not begin to bias matter power spectrum 
measurements until $k>70$ $h$/Mpc. As a result, we do not 
discuss any matter power spectra measurements for wavenumbers higher 
than $k=70$ $h$/Mpc. 

%A note on super-sample modes
We note that a complication arises 
when the matter power spectrum is measured 
within a sub-volume of a simulation 
due to an incomplete sampling of fluctuations. 
Modes with wavelengths larger than 
typical sub-volume scales, super-sample modes, observably impact 
how sub-sample modes evolve through nonlinear mode coupling. 
\citet{Hamilton:2005dx} originally pointed out 
the super-sample effect on the covariance 
of the power spectrum, where it was referred to as beat coupling. 
The super-sample effect vanishes when 
measuring the power spectrum throughout the entire volume 
of a simulation due to periodic boundary conditions.
\citet{Takahashi:2009bq} found that the measured power spectrum 
in comparison to the truth 
is biased low at wavenumbers corresponding 
to the sub-volume's side length, $k \sim 1/L_n$, but 
reduces progressively as the scales decrease, $k \gg 1/L_n$. 
Although, the super-sample effect does bias our matter power 
spectrum estimates, 
our results comparing one power spectrum with another will not be 
affected (i.e. the super-sample effect biases all of our estimates in the same way). 
The super-sample effect is less of an issue for our shear power spectrum 
analysis in section \ref{sec:weaklensing} (and observationally), where 
large scale modes along the line of sight are included.

\subsection{Effect of Super-Cluster Selection on the Matter Power Spectrum}
\label{sec:environment}

In this section we investigate how the overdensity 
of a region affects the matter power spectrum measured from that region's sub-volume. 
We examine this selection effect 
on the matter power spectra by comparing the matter power spectra measured from 
our random and super-cluster samples. 

\begin{figure}
 \includegraphics[width=\columnwidth]{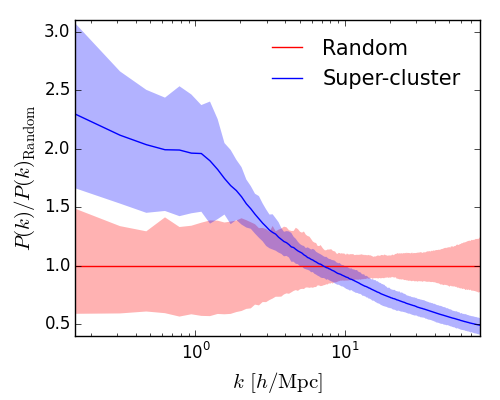}
 \caption{The median power spectra of the super-cluster and random 
 samples measured from the DMO re-simulations, both  
 divided by the median power spectrum of the 
 DMO random sample. 
 The ratio for the super-cluster and random sample is shown in blue and red 
 respectively, and the 
 shaded area surrounding the 
 curves represent the region between their 
 $16^{\mathrm{th}}$ and $84^{\mathrm{th}}$ percentiles. }
 \label{fig:3d_pk_overdmorand}
\end{figure}
In Fig. \ref{fig:3d_pk_overdmorand}, we plot the median 
matter power spectrum for the super-cluster DMO sub-volumes 
and that of the random DMO sub-volumes, both divided 
by the median matter power spectrum for the random DMO sub-volumes. 
Therefore, any deviation from unity is due to the effect of selection 
on the power spectrum.
The shaded area surrounding the blue and red curves 
represent the region between the super-cluster and random samples' 
$16^{\mathrm{th}}$ and $84^{\mathrm{th}}$ percentiles, 
which shows the scatter in the power spectra of the samples. 

On large scales,  $k \leq 1$ $h$/Mpc, the super-cluster 
sample has more than twice the power of the random sample on average.
As we move to smaller scales 
the excess power in the super-cluster sub-volumes 
begins to decline and beyond 
$k \sim 7$ $h$/Mpc, comparatively, 
the power of the 
random sample begins to dominate. 
On the smallest scales measured, $k \sim 70$ $h$/Mpc, 
the random sample now has double the power of the super-cluster sample. 

We note that the mean overdensity measured within the sub-volumes 
of our super-cluster and random samples 
are approximately one and zero 
respectively. 
The extra power seen in the super-cluster sample on large scales 
is expected. \citet{Chiang:2014oga} showed that a correlation 
exists between the matter power spectrum measured in 
a sub-volume 
and the overdensity of 
that sub-volume, with the more overdense sub-volumes having 
more power over the scales which they studied, $k \leq 1$ $h$/Mpc. 
To further investigate the effect of selection on the matter 
power spectrum we look at the individual sub-volumes of our 
two samples. 

\begin{figure}
 \includegraphics[width=\columnwidth]{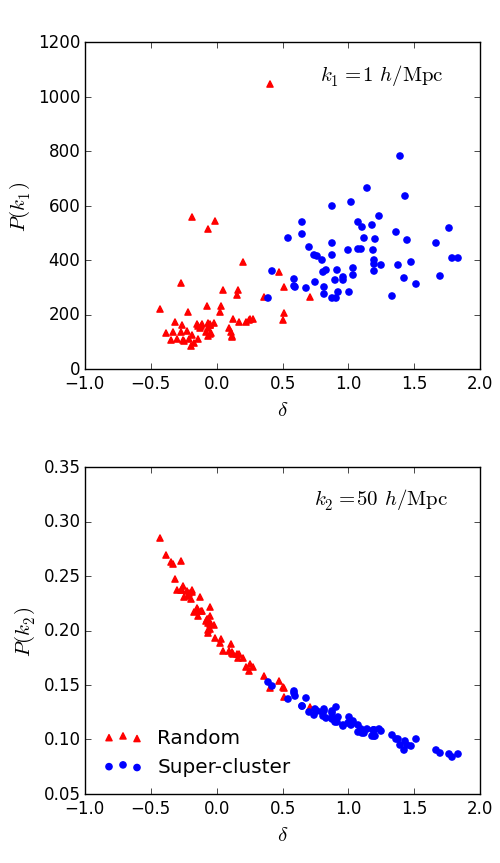}
 \caption{The dependence of the amplitude of the matter 
 power spectrum on the overdensity of the sub-volumes from which it was measured. 
 All 121 sub-volumes are shown, the super-cluster sample in blue and the random 
 sample in red. The top and bottom panels show the correlation 
 at scales corresponding to 
 $k_1=1 \: h/$Mpc and $k_2=50 \: h/$Mpc respectively. }
 \label{fig:pkdmo_overdensity1}
\end{figure}
In Fig. \ref{fig:pkdmo_overdensity1} we show, at two different scales, 
how the matter power spectrum measured in each DMO sub-volume varies 
with overdensity. 
In the top panel we plot the dependence of the amplitude of the 
matter power spectrum,
at $k_1=1 \: h/$Mpc, versus overdensity. 
In red are the random 
sub-volumes and in blue the super-cluster sub-volumes. 
With the exception of one outlier from the random sample, 
it can be seen that the power measured 
within a sub-volume increases with its overdensity, 
in agreement 
with \citet{Chiang:2014oga}. 

The equivalent plot for a smaller scale, 
$k_2=50 \: h/$Mpc, is shown in the bottom panel. We see a tighter correlation 
in the opposite direction, with more overdense sub-volumes having 
less power than their underdense counterparts. 
We found the same correlation at small scales in the HYDRO re-simulated 
sub-volumes and the original sub-volumes of the parent 
simulation for $k\ge10 \: h/$Mpc, which indicates 
this is not a resolution artefact. 
Halo assembly bias perhaps provides a plausible explanation for this effect. 
\citet{2008ApJ...687...12D} found that low mass halos 
in overdense regions are biased low, i.e. as the background density is raised, 
the number of low mass peaks are reduced as they are converted to high mass peaks. 
Another process that may contribute to the observed anti correlation 
is tidal stripping. 
In the more overdense regions these low mass halos are more likely 
to be in the presence of clusters, which could potentially 
strip these cluster galaxies of their mass.

\subsection{Effect of Baryons on the Matter Power Spectrum}
\label{sec:baryons}

In this section we investigate the effect of baryons on the matter 
power spectrum by comparing the matter power spectra 
measured from our HYDRO re-simulations with those 
of our DMO re-simulations. 

\begin{figure}
 \includegraphics[width=\columnwidth]{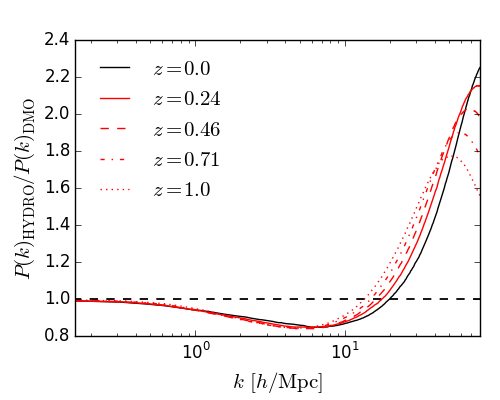}
 \caption{The median of the random sample's matter power spectra 
 measured from the HYDRO re-simulations divided by that of their 
 DMO counterparts. 
 The ratios measured at $z= 0.0, 0.24 ,0.46 ,0.71$ and $1.0$ are represented by the 
 solid black, solid red, dashed red, dash-dotted red and dotted red 
 curves respectively. }
 \label{fig:matterpowerspectrumbaryonsredshift}
\end{figure}
Plotted in Fig. \ref{fig:matterpowerspectrumbaryonsredshift} 
is the median of the matter power spectra 
measured from the HYDRO re-simulations 
divided by the matter power spectrum measured from their 
DMO counterparts, in our random sample. 
As the initial conditions of the HYDRO and DMO 
re-simulations are the same for a particular 
sub-volume, any deviation from unity results directly from 
the various effects of the baryonic physics present. 
The power spectra were measured at 
redshifts $z= 0.0, 0.24, 0.46 ,0.71$ and $1.0$ and 
their ratios are represented by the 
solid black, solid red, dashed red, dash-dotted red and 
dotted red curves respectively. 
There is no effect on the matter power spectrum 
due to baryons on the largest scales but they suppress 
the power on intermediate scales, $k\sim \: 0.3-10 \: h$/Mpc, and 
boost the power on small scales. 
Note this effect does not depend strongly on redshift. 

Our results are broadly consistent with previous work. 
\citet{2011MNRAS.415.3649V} compared 
three different hydrodynamical simulations 
to a reference dark matter only simulation, 
all of which were given the same initial conditions. 
Our simulations are most similar to their 
AGN model, which has 
baryonic processes that best resemble ours. 
They attributed the increase in power at the small scales 
to cooling baryons falling into potential wells, 
while at intermediate scales $k\sim \: 1-10 \: h$/Mpc
they showed the suppression of power was due 
to AGN feedback removing baryons 
from halos. 
On the largest scales, there is little to 
no effect on the power spectrum that results from baryons 
because the baryons are expected to trace 
of the dark matter on adequately large scales. 

\begin{figure}
 \includegraphics[width=\columnwidth]{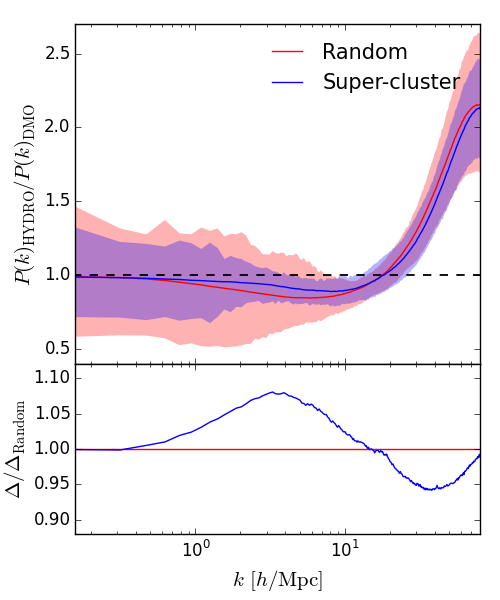}
 \caption{The top panel shows the 
 median of the matter power spectra measured in the 
 HYDRO re-simulations divided by that of their DMO counterparts, 
 for the random and super-cluster samples in red and blue respectively. 
 The shaded area surrounding the curves 
 represent the region between their 
 $16^{\mathrm{th}}$ and $84^{\mathrm{th}}$ percentiles.
 The bottom panel shows the super-cluster ratio divided by the random 
 ratio.}
 \label{fig:matterpowerspectrumbaryons}
\end{figure}
We can examine in more detail the effect baryons have 
on the matter power spectrum by comparing its effect 
on the super-cluster sample to that on the random sample. 
In the top panel of Fig. \ref{fig:matterpowerspectrumbaryons} we plot 
the median matter 
power spectrum measured in the HYDRO samples divided by that of the 
DMO samples. 
The red and blue curves represent 
the random and super-cluster HYDRO/DMO samples 
respectively and again the shaded area 
surrounding the curves 
represent the region between their 
$16^{\mathrm{th}}$ and $84^{\mathrm{th}}$ percentiles. 
On intermediate scales, $k\sim \: 1-10 \: h$/Mpc, 
the effect of baryons in the random sample suppresses the power of the HYDRO 
power spectrum to as little as 83\% of the DMO power spectrum. 
In comparison, the super-cluster sample's HYDRO power 
spectrum is suppressed 
to 88\% of the DMO power spectrum. 

The differences are shown more clearly in 
the bottom panel, where the HYDRO/DMO power spectrum of the super-cluster 
sample is divided by that of the random sample. 
The suppression of the matter power spectrum due to the effect 
of baryons can be seen to be stronger by several percent in the random sample. 
At small scales, beyond $k\sim \: 20 \: h$/Mpc, 
the boost in power due to baryons in the super-cluster sample is weaker 
than the boost in power due to baryons in the random sample. 
In short, at both intermediate and small scales, 
the effect of baryons on the matter power spectrum 
in the super-cluster sample is less severe than in 
the random sample. 

\begin{figure}
 \includegraphics[width=\columnwidth]{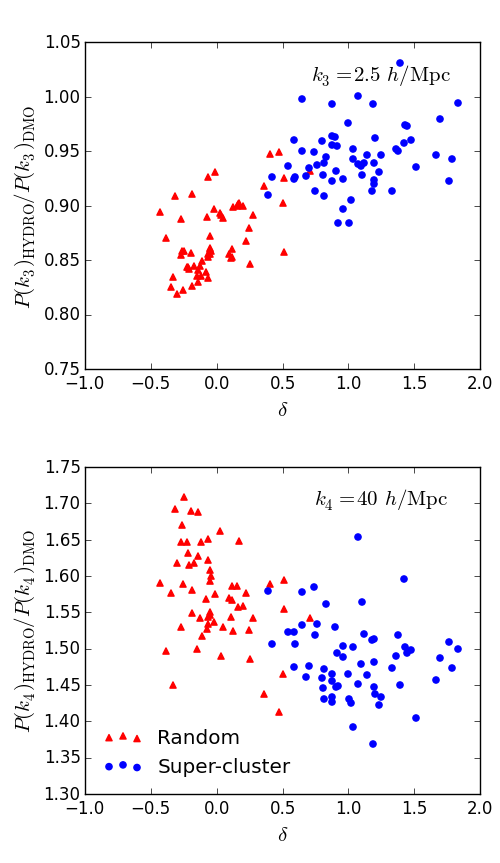}
 \caption{The dependence 
 of the HYDRO/DMO matter power spectrum ratio (effect of baryons) 
 on the overdensity of the sub-volumes from which they are measured. 
 All 121 sub-volumes are shown, the super-cluster sample in blue and the random 
 sample in red. The top and bottom panels show the correlation 
 at scales corresponding to 
 $k_3=2.5 \: h/$Mpc and $k_4=40 \: h/$Mpc respectively. }
 \label{fig:pkdmo_overdensity2}
\end{figure}
Similar to Fig. \ref{fig:pkdmo_overdensity1} we 
can also demonstrate how the effect of baryons is dependent 
on the overdensity of the sub-volume. 
In Fig. \ref{fig:pkdmo_overdensity2} we show 
how the HYDRO/DMO matter power spectra ratio in each 
sub-volume varies with its overdensity, at two different scales. 
The wavenumbers $k_3$ and $k_4$ were chosen to correspond 
to the peak and trough in the bottom panel of Fig. \ref{fig:matterpowerspectrumbaryons}. 
In red are the random sub-volumes 
and in blue the super-cluster sub-volumes. 
In the top panel we plot the dependence of this ratio on the overdensity, measured 
at intermediate scales, $k_3 = 2.5 \: h/$Mpc. 
Generally speaking, the higher the overdensity of a sub-volume 
the less the power is suppressed by baryons. 
A plausible explanation is that on 
these scales, 
the AGN feedback 
becomes less efficient at removing 
gas from the deeper potentials. 
As a result, more gas stays in the haloes and it is likely 
this is the cause of the slight discrepancy seen in Fig. \ref{fig:fbarm500}. 

The bottom panel shows the equivalent plot at small scales 
$k_4=40 \: h/$Mpc, where we can say in general that the 
increase in power due to cooling baryons 
falling into potential wells is less severe for the more overdense regions. 
This is likely related to the presence of more massive haloes but may 
also be due to the longer cooling time of the hotter gas associated with 
these haloes, as well as the increased AGN feedback.

\subsection{Comparing the Effect of Baryons with Super-Cluster Selection}
\label{sec:environmentbaryons}

\begin{figure}
 \includegraphics[width=\columnwidth]{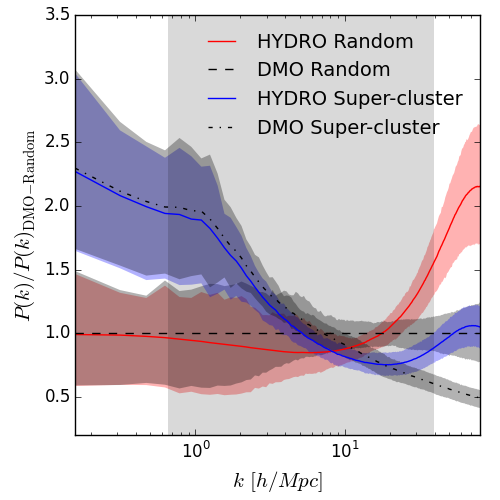}
 \caption{The ratios of the median matter power spectrum measured from a sample and 
the median matter power spectrum measured in the 
random sample's DMO re-simulations. The ratio 
for the HYDRO supercluster sample, DMO supercluster sample, HYDRO random sample and 
DMO random sample are shown in solid blue, black dash-dotted, solid red and black dashed 
respectively. The shaded area surrounding the curves 
represent, the region between their 
$16^{\mathrm{th}}$ and $84^{\mathrm{th}}$ percentiles.
The grey shaded region represents the scales which Super-CLASS could probe, 
assuming a one square degree field of view and a source density sufficient 
to bin galaxies in pixels of one square arcmin.}
 \label{fig:matterpowerspectrumenvironment}
\end{figure}
In the previous two sub-sections we have looked at how the matter power 
spectrum is affected by baryons and selection separately. 
In this section we will compare the two effects, specifically to determine 
which of the two is more important. 

Plotted in Fig. \ref{fig:matterpowerspectrumenvironment} 
are the median power spectra measured in the HYDRO super-cluster, 
HYDRO random, DMO super-cluster and DMO 
random re-simulations divided by 
the median power spectrum of the DMO random re-simulations, 
shown by the blue, red, black dash-dotted and 
black dashed curves respectively. 
Dividing the median power spectra by that of the DMO random re-simulations, 
allows us to compare the effect of baryons and 
selection simultaneously. 
Again the shaded area 
surrounding the curves 
represent the region between their 
$16^{\mathrm{th}}$ and $84^{\mathrm{th}}$ percentiles. 

On the largest scales, as we showed previously, there is little to 
no effect on the power spectrum that results from baryons, 
as the distribution of baryons trace that 
of the dark matter. 
In contrast, the power of the super-cluster sample is over double that 
of the random sample. 
The super-cluster selection effect 
boosts the matter power spectrum and is dominant, 
down to scales $k\sim \: 5 \: h$/Mpc, over the 
opposing effect of the baryons which suppresses the power. 

The blue curve which is the median power spectrum measured 
from the HYDRO re-simulations of the super-cluster sample, shows 
which of the two effects is dominant at scales where they oppose each other. 
It can be seen that on scales between $k\sim \: 5-7 \: h$/Mpc that the 
baryonic effect starts to dominate. 
On scales between $k\sim \: 7-20 \: h$/Mpc baryons continue to suppress 
the power, while the selection effect 
of the super-cluster sample 
now switches sign and compliments the baryons by further 
suppressing the power. On 
these scales both baryonic and selection effects are similar. 

On smaller scales, $k\geq \: 20 \: h$/Mpc, the effect of baryons 
is to boost the power and the suppression of 
power due to the selection effect begins to dominate 
again, until $k\sim \: 50 \: h$/Mpc. On scales 
smaller than this the boost in power 
from baryons dominates. 
The shaded grey region shows the scales which we will 
be able to probe with a weak lensing survey like Super-CLASS. 
Except for a small range of scales, the selection effect of 
the super-cluster sample dominates over the effect of baryons. 

To summarise, the amount to 
which the selection effect and baryons affect 
the matter power spectrum on small scales is similar, but they cancel 
each other out to a degree. 
On intermediate scales, which 
current weak lensing studies can probe, the dominant effect is selection. 
In addition, we showed in 
Fig. \ref{fig:matterpowerspectrumbaryonsredshift} 
that the effect of baryons on the matter power spectrum does not 
vary significantly with redshift. 
For these reasons, 
in the next section we will ignore baryonic effects and 
look at how the more important selection effect of 
targeting a super-cluster region biases 
the primary 2-point statistic used in weak lensing surveys, 
the shear power spectrum.

\section{Weak Lensing Shear Power Spectrum}
\label{sec:weaklensing}

In this section we describe the method we implemented 
to generate convergence maps along lightcones that intersect 
our random and super-cluster sub-volumes on the way through the 
parent simulation. We then examine the effect 
of selection, i.e. the effect 
of targeting a super-cluster field as opposed to a random field, on 
the shear power spectrum by comparing our random 
and super-cluster convergence maps, before presenting some 
forecasts for Super-CLASS like weak lensing surveys. 

The distortion of a source galaxy, in an image at 
position $\boldsymbol{\theta}=(\theta_1, \theta_2 )$, due to the effect of weak 
gravitational lensing can be described by two fields. 
The scalar convergence field $\kappa(\boldsymbol{\theta})$ describes 
the change in apparent size of the source galaxy, 
while the complex valued shear field $\gamma(\boldsymbol{\theta})$ 
describes the compression and stretching 
of the source galaxy. 
Both of these can be expressed in terms of a lensing 
potential field $\psi(\boldsymbol{\theta})$ as follows,
\begin{equation}
 \label{shear}
 \gamma(\boldsymbol{\theta})=\gamma_{1}(\boldsymbol{\theta}) + i\gamma_{2}(\boldsymbol{\theta}) = 
 \frac{1}{2} \left (\partial_1^2 - \partial_2^2  \right )\psi(\boldsymbol{\theta}) + i \partial_1 \partial_2 \psi(\boldsymbol{\theta}),
\end{equation}
\begin{equation}
 \label{convergence}
 \kappa(\boldsymbol{\theta}) = \frac{1}{2} \left (\partial_1^2 + \partial_2^2  \right ) \psi(\boldsymbol{\theta}),
\end{equation}
where partial derivatives $\partial_1$ and $\partial_2$ are 
with respect to $\theta_1$ and $\theta_2$ respectively, and 
$\gamma_{1}$ and $\gamma_{2}$ are the two components of 
the shear field. 
Additionally, the convergence field can be related to the projected 
surface mass density $\Sigma(\boldsymbol{\theta})$ by 
\begin{equation}
 \label{surfacemassdensity}
 \kappa(\boldsymbol{\theta}) = \frac{\Sigma(\boldsymbol{\theta})}{\Sigma_{\mathrm{crit}}},
\end{equation}
where the critical surface mass density is defined by
\begin{equation}
 \label{criticalsurfacemassdensity}
 \Sigma_{\mathrm{crit}} = \frac{c^2}{4\pi G} \frac{D_{\mathrm{s}}}{D_{\mathrm{l}} D_{\mathrm{ls}}},
\end{equation}
where 
$D_{\mathrm{ls}}$, $D_{\mathrm{s}}$ and $D_{\mathrm{l}}$ are the angular diameter 
distances from the lens to the source, the observer to the source, 
and the observer to the lens respectively. 

\subsection{Convergence and Shear Maps}
\label{sec:lightcones}
In section \ref{sec:matterpowerspectrum} we showed 
the effect due to measuring the 
matter power spectrum in a super-cluster sub-volume is 
more important than the effect of baryonic physics. 
Additionally, we showed the effect of baryons on the matter 
power spectrum does not depend greatly on redshift nor 
the type of sub-volume 
in which the power spectrum was measured. 
This leads us to believe the effect of baryons on the matter power 
spectrum is fairly consistent across a range 
of different sub-volumes at different redshifts, 
up to a minimum of $ z = 1 $. 
However, given the large effect on the matter power spectrum 
due to selection, we would also like to examine how much 
selection biases the shear power spectrum. We went about this 
by generating convergence maps that targeted our random 
and super-cluster sub-volumes in the parent simulation. 

The convergence in angular pixels was calculated
by generating lightcones through our parent simulation, using 
the method developed in the $\textsc{sunglass}$ pipeline
\citep[for more details see,][]{2011MNRAS.414.2235K}, 
\begin{equation}
 \label{pixelconvergence}
 \kappa_{\mathrm{p}}(r_s)=\sum _k \frac{K(r_k,r_s)}{\Delta \Omega_p \bar{n}(r_k) r_k^2}
- \int_{0}^{r_s}\mathrm{d}rK(r,r_s),
\end{equation}
where $\Delta \Omega_p$ is the pixel area, $\bar{n}$ is the simulation's comoving 
particle number density, and $r_s$ and $r_k$ are the comoving radial distance 
to the source plane and position of the $k^{\mathrm{th}}$ particle in the lightcone, respectively. 
The scaled lensing kernel is defined as follows, 
\begin{equation}
 \label{lensingkernel}
 K(r,r_s)=\frac{3 H_0^2 \Omega_{\mathrm{m}}}{2c^2} \frac{(r_s-r)r}{r_s a(r)}.
\end{equation}
This method exploits the Born approximation to 
perform a line of sight integration 
along the unperturbed ray path. 
This simplification is justified as the effects of 
lens-lens coupling and higher order 
corrections of the Born approximation 
on the convergence power spectrum have been 
shown to be insignificant on relevant scales \citep[see ][]{2009A&A...499...31H}. 

We generated lightcones orientated such that they 
intersected our super-cluster and random sub-volumes 
at the snapshot corresponding to $z=0.24$, to create 
one square degree convergence maps with a single 
source plane at $z = 1$. 
This was chosen to match the median redshift 
of the source galaxies estimated in the SKA forecast by \citet{Bonaldi:2016lbd}. 
As Super-CLASS has the same frequency, depth and resolution as SKA 
it will likely have galaxies with a very similar redshift distribution. 
The angular size of our convergence maps were chosen 
to match that of the Super-CLASS field. 
The parent simulation volume was split into equal length sections 
of comoving distance 400 Mpc along the line of sight, 
in order to include the evolution of structure. 
The first 400 Mpc section used was from the 
snapshot corresponding to $z=0.05$, 
this was set up so that the time it takes light to travel from 
the centre of the section to the start of the lightcone is equal to how much 
time has passed from $z=0.05$ to now. 
All particles that fell 
within the lightcone's volume were added to the line of sight integration, 
before adding the next section of the lightcone from the next snapshot. 
This continued until the snapshot that corresponds to $z=0.85$ where 
the lightcone reached the end of the parent simulation volume at 3200 Mpc. 
As the parent simulation has continuous boundary conditions the next section 
was naturally taken from the opposite side from the final snapshot corresponding 
to redshift $ z = 1.0 $. 
The particles were binned into a 600$^2$ grid of pixels, with each 
square pixel having an area $A = 0.01$ arcmin$^2$. 

Once the convergence maps were generated, we calculated 
shear maps in Fourier space 
using the following relations, 
\begin{equation}
 \label{gamma1}
 \hat{\gamma}_1(\mathbf{\ell})=\frac{\ell_1^2-\ell_2^2}{\ell_1^2+\ell_2^2} \hat{\kappa}(\mathbf{\ell}),
\end{equation}
\begin{equation}
 \label{gamma2}
 \hat{\gamma}_2(\mathbf{\ell})=\frac{2 \ell_1 \ell_2}{\ell_1^2+\ell_2^2} \hat{\kappa}(\mathbf{\ell}),
\end{equation}
where $\ell_1$ and $\ell_2$ are the Fourier 
conjugates of $\theta_1$ and $\theta_2$, $\hat{\kappa}$ is the Fourier 
transform of the convergence, and $\hat{\gamma}_1$ and $\hat{\gamma}_2$ 
are the Fourier transforms of the two components of the shear. 

\begin{figure*}
 \includegraphics[width=\textwidth]{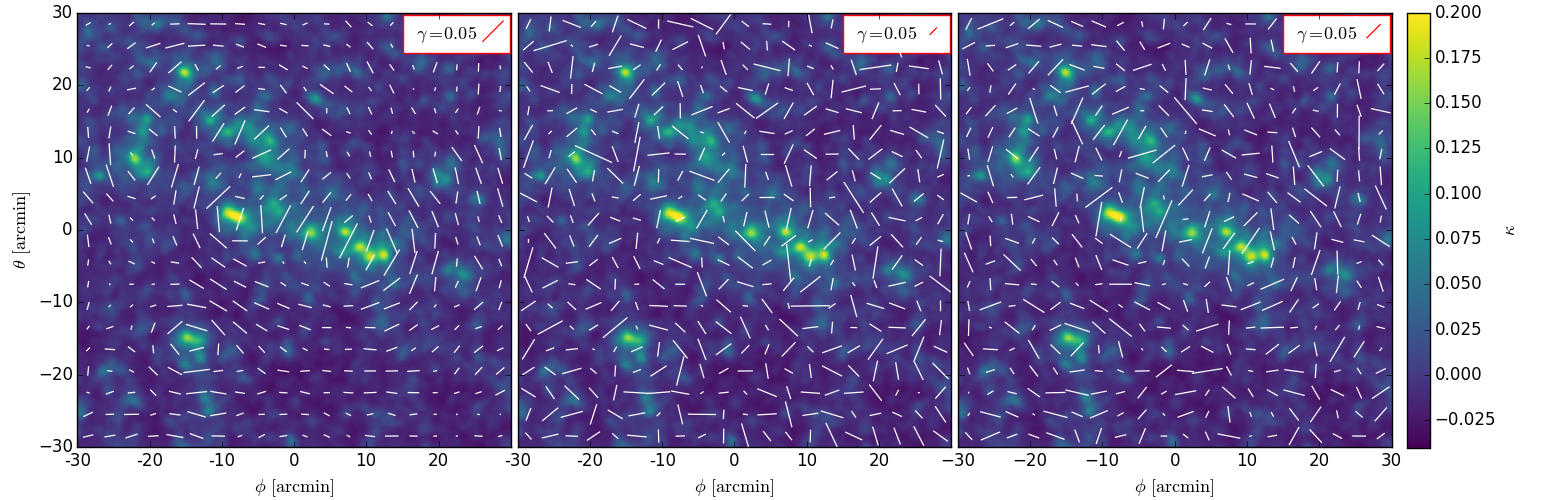}
 \caption{All three panels 
 show one square degree, convergence and shear maps 
 for one member of our super-cluster sample. 
 The background and colour bar show the convergence map, while 
 the white ticks show the shear field. 
 In the left panel the shear ticks represent the shear field before adding noise, while 
 the middle and right panels show the shear fields after adding noise 
 typical of current radio and optical weak lensing surveys, respectively. }
 \label{fig:lightcone_convergence_shear_map}
\end{figure*}

The left panel of Fig. \ref{fig:lightcone_convergence_shear_map} 
shows one example of a convergence and shear map 
taken from our super-cluster sample. 
The integrated convergence field up to $z=1$ 
is shown by the background colour, where the yellow and dark blue 
regions indicate overdense and underdense areas respectively. 
The shear field is shown by the white ticks 
and can be seen to 
tangentially trace the overdense areas 
as expected.

\subsection{Effect of Super-Cluster Selection on the Convergence Power Spectrum}
\label{sec:environmentweaklensing}
\begin{figure}
 \includegraphics[width=\columnwidth]{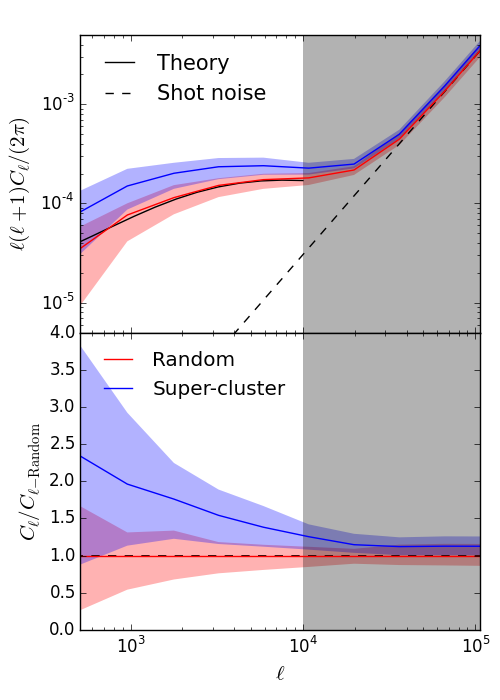}
 \caption{The top panel shows the convergence power spectra of the random 
 and supercluster samples, in red and blue respectively. 
 The shaded areas 
 surrounding the two curves 
 represent the region between their 
 $16^{\mathrm{th}}$ and $84^{\mathrm{th}}$ percentiles. 
 The smooth black curve is the predicted theoretical 
 convergence power spectrum from $\textsc{nicaea}$, 
 while the diagonal black dashed 
 line shows the shot noise term. 
 In the bottom panel we plot the convergence power spectra of the random 
 and supercluster samples divided by the median convergence power spectrum 
 of the random sample. 
 The grey shaded region in both panels show the 
 scales for which the measured power spectra 
 begin to be dominated by shot noise. }
 \label{fig:lightcone_Cl_l}
\end{figure}

In this section we present our results for 
the effect of super-cluster selection on the convergence power spectrum. 
The convergence and shear power spectra, radially 
binned in $\ell=(\ell_1^2+\ell_2^2)^{1/2}$, can be determined from the 
convergence and shear fields respectively using 
\begin{equation}
 \label{clk}
 C_{\ell}^{\kappa \kappa}=\left (\frac{L}{N}  \right )^2 \sum_{\ell \, \mathrm{in \, shell}} 
 \left \langle \left | \hat{\kappa} \right |^2 \right \rangle,
\end{equation}
\begin{equation}
 \label{clgamma}
 C_{\ell}^{\gamma \gamma}=\left (\frac{L}{N}  \right )^2 \sum_{\ell \, \mathrm{in \, shell}} 
 \left \langle \left | \hat{\gamma}_1 \right |^2 \right \rangle+
 \left \langle \left | \hat{\gamma}_2 \right |^2 \right \rangle,
\end{equation}
where $L$ is the angular size of the image field in radians and $N$ is 
the number of pixels in the image. 
Here 
we use angle brackets to refer to the mean of a variable. 
We calculated the convergence power spectrum 
for every realization in the two samples. 
When measuring the power spectrum 
from a simulation with finite particles we 
measure $C_{\ell}^{\mathrm{m}}=C_{\ell}^{\kappa \kappa}+C_{\ell}^{SN}$, 
the desired power spectrum with an additional shot noise 
component which can be modelled as 
\begin{equation}
 \label{clSN}
 C_{\ell}^{SN}=\left (\frac{3 H_0^2 \Omega_\mathrm{m}}{2 c^2}  \right )^2\int_{0}^{r_s} 
 \frac{(r_s-r)^2}{\bar{n}(r)r^2_s a(r)^2}\mathrm{d}r.
\end{equation}
The mean 3D number density of particles in the simulation $\bar{n}(r)$, will 
be constant in comoving coordinates. 

The top panel of Fig. \ref{fig:lightcone_Cl_l} shows  
the median convergence power spectra 
for our random and super-cluster samples in red and 
blue respectively, where
again the shaded area 
surrounding the curves 
represent the region between their 
$16^{\mathrm{th}}$ and $84^{\mathrm{th}}$ percentiles. 
We used the publicly available code 
$\textsc{nicaea}$\footnote{http://www.cosmostat.org/software/nicaea/} \citep{Kilbinger:2008gk} 
to predict the theoretical convergence power spectrum, 
shown by the black curve, assuming 
the revised $\textsc{halofit}$ \citep{Takahashi:2012em} model for the 
input non-linear matter power spectrum. 
The shot noise power spectrum is shown by the 
dotted black line. 
The median convergence power spectrum of 
the random sample shows good agreement with the theoretical power spectrum on 
the scales for which it is predicted. On smaller scales, shaded in grey 
($\ell>10^4$), the measured power spectra 
begins to be dominated by shot noise. 

The bottom panel shows the power spectra of the random and super-cluster 
samples divided by the median power spectrum of the random sample. 
This lets us see the effect of selection on the convergence power spectra 
over a range of scales. 
We see that on the largest scales 
the median convergence power spectrum for the super-cluster sample 
is over twice as large as that for the random sample. 
On smaller scales, $\ell \sim 10^4$, the super-cluster sample 
still has approximately $50$ percent 
more power than the random sample. 
Although we cannot directly compare the matter power spectrum results with these, 
we can see similarities here with the results of Fig. \ref{fig:3d_pk_overdmorand}. 
The boost in power at high $\ell$ values is likely due to contributions 
from lower $k$ modes at higher redshifts.

Here we only show the results for the convergence field 
with a source plane at $z=1$. 
However, the same method was used to generate 
convergence fields at $20$ source redshift planes separated by $\Delta z=0.1$, 
from $z=0.1$ to $z=2$. The median convergence power spectrum for the 
super-cluster sample is larger than that of the random sample in all slices 
with redshifts higher than that of super-cluster (i.e. $z\ge0.3$). At the largest scale 
the median convergence power spectrum for the super-cluster sample range between 1.7 times 
and 2.7 times larger than that of random sample, for $z=2.0$ and $z=0.5$ respectively.

The minimum $\ell$ value is limited by the size of our field of view 
and can be estimated using $\ell_{\mathrm{min}} \sim \pi/\theta_{\mathrm{FOV}}=180$, 
where $\theta_{\mathrm{FOV}} =1$ degree is the angular size of the field of view. 
In lensing surveys, the maximum $\ell$ value is limited 
by the angular size of the pixels $\theta_{\mathrm{pix}}$ in which you are able to bin 
a sufficient number of galaxies, i.e. $\ell_{\mathrm{max}} \sim \pi / \theta_{\mathrm{pix}}$. 
In the Super-CLASS survey we expect $\theta_{\mathrm{pix}}$ to be around one arcmin, 
which means $\ell_{\mathrm{max}} \sim 10^4$. 
Therefore, the approximate range of scales 
most relevant to this work is, $180 < \ell < 10^4$.

\subsection{Forecasts for the Super-CLASS Survey}
\label{sec:forecasts}
In the previous sub-section we found that the amplitude of the convergence 
power spectrum measured from a lightcone which intersects 
a super-cluster sub-volume is larger 
on all scales than the equivalent for a random sub-volume. 
The motivation for this study was to 
use simulations to forecast the constraining power of the 
Super-CLASS survey. 
In this section we add noise to 
our shear fields, in order to 
estimate 
how well we will be able to measure the shear power 
spectrum from the Super-CLASS radio data. 

%Systematics
There are other systematics and sources of bias
that can influence shear power spectrum constraints, e.g. galaxy 
shape and redshift measurements. In addition, certain systematics 
will affect cosmic shear measurements differently in radio and 
optical weak lensing surveys, e.g. the stability of the telescope beam 
in the radio band is advantageous as it is almost unaffected by seeing. 
The main physical systematic is a correlation in the ellipticities of galaxies 
prior to lensing, i.e. intrinsic alignments. However, nulling techniques and methods 
in which galaxy pairs are weighted based on 
there spatial separation can be used to reduce the impact of intrinsic 
alignments on the measured power. 
These methods require the redshift distribution of the 
source galaxies in the survey, which Super-CLASS will include. 
In addition, redshift measurements could also be used 
to ensure galaxies that are physically associated with the lens are 
excluded from the source catalog, in order 
to reduce the boost factor \citep{2015ApJ...806....1M}, which 
dilutes the signal. 
We will not investigate the effects of these systematics here as they will 
be addressed in forthcoming Super-CLASS papers. 

%how to add noise
When observational data collected from weak lensing surveys 
is used to investigate cosmic shear, 
shape measurements of the galaxies are used to 
estimate the shear. 
The estimated shear is determined 
by binning a finite number of galaxies into pixels, 
before taking the average of their ellipticities.
To simulate the associated shape noise we added Gaussian noise to both 
components of the shear in real space as follows, 
\begin{equation}
 \label{gamma1noise}
 \gamma_1^n(\theta)= \gamma_1(\theta) + N_1(\theta),
\end{equation}
\begin{equation}
 \label{gamma2noise}
 \gamma_2^n(\theta)= \gamma_2(\theta) + N_2(\theta),
\end{equation}
where $\gamma_1^n(\theta)$ and $\gamma_2^n$ are the 
noisy shear components, and $N_1$ and $N_2$ are the 
added noise fields. 
Both noise fields were generated with zero mean and a standard 
deviation, $\sigma_n=\sigma_{\epsilon}/N_p^{1/2}$, where $N_p=N_a A$ is the number 
of galaxies per pixel, $N_a$ is the number 
of galaxies per arcmin$^2$, and $\sigma_{\epsilon}$ is the rms of the typical 
intrinsic ellipticity. We have taken $\sigma_{\epsilon}=0.3/\sqrt{2}$ which is 
typical of current ground based surveys.

The middle and right panel of Fig. \ref{fig:lightcone_convergence_shear_map} 
show two shear fields with added noise. 
We used a 
source density that is expected for the 
Super-CLASS radio survey, $N_a^R=1.5$ galaxies/arcmin$^2$, 
for the middle panel. For the right panel we used 
a source density that is typical of high quality ground based optical 
surveys, $N_a^O=10$ galaxies/arcmin$^2$. 
The integrated convergence map up to $ z = 1 $ 
is shown by the background and colour bar, and is the 
same in all three panels. 
The impact of adding noise is largest for the radio scenario, 
the middle panel. The shear ticks no longer tangentially trace 
the high convergence regions perfectly, as they do in the left panel. 
In the optical scenario, the right panel, the shear ticks trace 
the high convergence regions tangentially fairly well. It is 
in the low convergence regions where the effect of noise is strongest. 

%how to calculate the e and b mode decomposition of the shear field
To investigate the effect of noise quantitatively 
we examine the shear power spectra recovered from the simulations. 
The shear is a spin-2 field and can be decomposed into gradient ($E$) 
and curl ($B$) components. The scalar potential, from which the shear field arises, 
only produces $E$ modes. Systematic effects and shape noise present in weak 
lensing surveys produce both $E$ and $B$ modes. 
As a result, the $B$ modes can be utilised 
to test for these effects. 
As an example, if we were to take the $E$ and $B$ components 
of our shear field before adding noise, we would find 
no $B$ component and the $E$ component 
would be the original convergence field. 
The result is not as trivial for our noisy shear field. 
In the Fourier domain we can determine the 
$E$ component of the shear field $\hat{\gamma}^E(\mathbf{\ell})$ and the 
$B$ component of the shear field $\hat{\gamma}^B(\mathbf{\ell})$ 
using the following relations, 
\begin{equation}
 \label{Emode}
 \hat{\gamma}^E(\mathbf{\ell}) = \frac{\ell_1^2-\ell_2^2}{\ell_1^2+\ell_2^2} \hat{\gamma}_1^n(\mathbf{\ell})+ 
 \frac{2 \ell_1 \ell_2}{\ell_1^2+\ell_2^2} \hat{\gamma}_2^n(\mathbf{\ell}),
\end{equation}
\begin{equation}
 \label{Bmode}
 \hat{\gamma}^B(\mathbf{\ell}) = \frac{2 \ell_1 \ell_2}{\ell_1^2+\ell_2^2}  \hat{\gamma}_1^n(\mathbf{\ell})- 
 \frac{\ell_1^2-\ell_2^2}{\ell_1^2+\ell_2^2} \hat{\gamma}_2^n(\mathbf{\ell}).
\end{equation}
The power spectra of 
the $E$ and $B$ modes, radially 
binned in $\ell=(\ell_1^2+\ell_2^2)^{1/2}$, can be calculated using 
\begin{equation}
 \label{clEmode}
 C_{\ell}^{EE}=\left (\frac{L}{N}  \right )^2 \sum_{\ell \, \mathrm{in \, shell}} 
 \left \langle \left | \hat{\gamma}^E \right |^2 \right \rangle,
\end{equation}
\begin{equation}
 \label{clBmode}
 C_{\ell}^{BB}=\left (\frac{L}{N}  \right )^2 \sum_{\ell \, \mathrm{in \, shell}} 
 \left \langle \left | \hat{\gamma}^B \right |^2 \right \rangle,
\end{equation}
where $L$ is the angular size of the image field in radians and $N$ is 
the number of pixels in the image. 

\begin{figure}
 \includegraphics[width=\columnwidth]{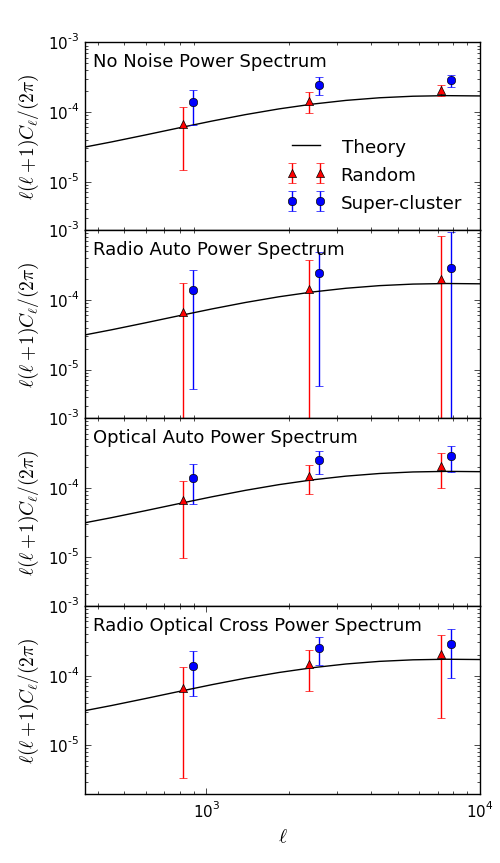}
 \caption{Mean shear power spectra of the random 
 and super-cluster samples are shown in red and blue respectively. 
 The top panel shows the power spectra before adding noise. The 
 second , third and fourth panel from the top show 
 the radio auto power spectrum, the optical auto power spectrum 
 and the radio optical cross power spectrum respectively. 
 The method used to determine the error bars is described in the text.}
 \label{fig:lightcone_signal_test_1gal}
\end{figure}

The shear field generated from our simulations can only produce $E$ modes. 
Therefore, the $E$ mode power spectrum contains a signal plus the added noise component. 
The $B$ mode power spectrum traces the noise component and 
can be used to estimate the noise bias. 
We subtracted the mean $B$ mode power spectrum from 
the $E$ mode power spectra 
$\widehat{C}_l^{EE} =  \widetilde{C}_l^{EE} - \left \langle \widetilde{C}_l^{BB} \right \rangle$, 
for both the random and super-cluster samples, 
and their mean values are plotted 
in red and blue respectively in Fig. \ref{fig:lightcone_signal_test_1gal}. 
The shear power spectrum 
with no noise is shown in the top panel. 

The Super-CLASS survey is expected to make shear estimates 
using overlapping radio and optical observations. As a result, 
we forecast the constraining power of the Super-CLASS 
survey for both optical and radio galaxy source densities. 
In the second panel down we show the 
radio auto power spectrum, for which we have 
used the source density expected for Super-CLASS, 
$N_a^R=1.5$ galaxies/arcmin$^2$. 
For the optical auto 
power spectrum plotted in the third panel down, we used a 
source density typical of high quality ground based optical 
surveys, $N_a^O=10$ galaxies/arcmin$^2$. The only 
difference between these two auto power spectra are 
the source densities (noise levels) chosen. 
Note that we are ignoring the possibility that the source density in a super-cluster region 
could differ to that in a random region. It is likely that fewer source galaxies would 
be observed in a super-cluster region due to obscuration by the foreground clusters. 
On the other hand the effect of magnification, which would allow fainter sources 
to be observed, may counteract this to some degree. 
For this analysis we do not consider these issues and assume the same 
source galaxy densities for both the random 
and super-cluster samples. 
Plotted in the bottom panel is the radio optical cross 
power spectrum. The solid black curve plotted 
in all four panels is the theoretical 
prediction from $\textsc{nicaea}$ again. 
We calculated the error bars using 
\begin{equation}
 \label{sigma_clE}
\sigma_l = \sqrt{\left \langle (\widehat{C}_l^{EE})^2 \right \rangle 
- \left \langle \widehat{C}_l^{EE} \right \rangle ^2}, 
\end{equation}
where the angle brackets indicate the average taken over our suite 
of simulations (i.e. our super-cluster or random sample).

\begin{table*}
 \caption{Median signal to noise and detection significance values of the random and super-cluster 
 samples for varying source densities in the absence of systematic effects. The error bars shown represent the 
$16^{\mathrm{th}}$ and $84^{\mathrm{th}}$ percentiles of the respective sample's 
distribution. }
 \label{tab:detection}
 \begin{tabular}{l c c c c c}
  \hline
  Power Spectrum & Source Density & Random & Random & Super-cluster & Super-cluster \\
  & galaxies/armin$^2$ & $S/N$ & $D$ & $S/N$ & $D$\\
  \hline
  No Noise & $\infty$ & $6.5^{+1.2}_{-1.4}$ & $\infty$ & $6.3^{+1.5}_{-1.3}$ & $\infty$\\[0.08cm]
  Radio Auto & 1.5 & $1.7^{+0.9}_{-0.7}$ & $2.0^{+1.0}_{-0.9}$ & $2.0^{+1.1}_{-0.8}$ & $2.7^{+1.5}_{-1.2}$\\[0.08cm]
  Optical Auto & 10 & $3.3^{+1.1}_{-1.0}$ & $7.4^{+3.2}_{-2.3}$ & $4.1^{+1.4}_{-1.1}$ & $13.7^{+8.0}_{-4.7}$\\[0.08cm]
  Radio Optical Cross & 1.5, 10 & $2.5^{+1.0}_{-0.8}$ & $4.4^{+2.0}_{-1.6}$ & $3.3^{+1.3}_{-1.0}$ & $7.6^{+4.6}_{-2.7}$\\
  \hline
 \end{tabular}
\end{table*}
The signal to noise $S/N$, 
the precision with which the 
cosmic shear signal can be measured, was determined for the random and 
super-cluster samples using 
\begin{equation}
 \label{signalnoise}
 S/N = \sqrt{\sum_{l}\left ( \frac{ \widehat{C}_l^{EE}}{\sigma_l} \right )^2}.
\end{equation}
Additionally, the detection significance $D$, 
the significance with which a non 
zero signal can be detected, 
was calculated as follows, 
\begin{equation}
 \label{detection}
 D = \sqrt{\sum_{l}\left ( \frac{ \widehat{C}_l^{EE}}{{\sigma}'_l} \right )^2}, 
\end{equation}
where 
\begin{equation}
 \label{sigma_clB}
 {\sigma}'_l = \sqrt{\left \langle (\widehat{C}_l^{BB})^2 \right \rangle 
- \left \langle \widehat{C}_l^{BB} \right \rangle ^2},
\end{equation}
and $ \widehat{C}_l^{BB} =  \widetilde{C}_l^{BB} - \left \langle \widetilde{C}_l^{BB} \right \rangle$. 
The signal to noise 
depends on both the 
shape noise and the sample variance, 
while the detection significance 
depends solely on the former. 
The signal to noise and detection significance for all four scenarios 
are shown in Table \ref{tab:detection}, 
the errors represent the 
$16^{\mathrm{th}}$ and $84^{\mathrm{th}}$ percentiles of the 
distribution. Note all values displayed in Table \ref{tab:detection} 
are the mean over one thousand realizations of the noise 
and do not take into account any systematic effects.

%No noise
The error bars in the first scenario, in which no shape noise was added to the shear field, 
for both the signal to noise values and the shear 
power spectrum are entirely due to sample variance. 
This provides us with an idealised scenario, which we can 
compare the effect of varying shape noise to. 
We found that the signal to noise and 
detection significance in the three noise scenarios 
are higher for the super-cluster sample than they are for the random sample. 
This is due to the amplitude of the convergence power spectrum 
of the super-cluster sample being larger than that of the random sample, 
as was shown in Fig. \ref{fig:lightcone_Cl_l}. 

%Radio power spectra
We should reiterate that the following results ignore all systematic effects. 
The simulated radio data has a detection significance of 
$2.7^{+1.5}_{-1.2}$ for the super-cluster sample, which 
is likely high enough to make a cosmic shear detection. 
The difference between surveying a super-cluster region as opposed 
to a random region, 
could make a cosmic shear detection possible, assuming 
the source galaxy density 
predicted for Super-CLASS is achieved. 
This was indeed the reasoning for an overdense 
super-cluster region of the sky to be chosen as 
the target of the 
Super-CLASS field. 

%Optical power spectra
As is to be expected, 
the third panel of the 
figure and row of the table show a more promising scenario for 
a source density typical 
of ground based optical surveys. 
In this case both the signal to noise and detection 
significance are sufficient for a cosmic shear detection, even in the random sample. 

%Cross power spectra
In the bottom panel and row, the optical radio cross power spectrum 
is seen to be a significant improvement on the 
radio auto power spectrum. 
The overlapping radio and optical observations 
when cross correlated produce a detection significance of 
$7.6^{+4.6}_{-2.7}$, which is a strong cosmic shear detection. 

\section{Summary and Conclusions}\label{sec:conclusion}
The radio weak lensing survey Super-CLASS 
aims to measure the cosmic shear signal by targeting an overdense  
region of the sky that contains five Abell clusters. 
Two issues that are relevant to the detection of this 
signal are the effect of selection and the 
impact of baryons on the matter distribution. 
We have investigated both issues in this paper 
by identifying a sample of super-cluster and random sub-volumes 
from a large dark matter only (parent) simulation. These sub-volumes 
were re-simulated, using zoom techniques, at a higher 
resolution and with full gas physics. 
This enabled us to examine the 
effect of baryons on the matter power spectrum in these sub-volumes 
along with the aforementioned selection effect. 
We also generated shear and convergence maps using the line 
of sight integration technique, which intercept the random 
and supercluster regions at $ z = 0.24 $. These were then used 
to study the difference between the 
shear power spectra measured by 
a weak lensing survey that targets a random patch of the sky 
as opposed to one containing a super-cluster. 
Our key results are summarised as follows: 

\begin{itemize}
\item
On large scales ($k \leq 1 \: h/$Mpc) the 
matter power spectrum measured from our super-cluster 
sample has at least twice as much power 
as that measured from the random sample 
(Fig. \ref{fig:3d_pk_overdmorand}). 
Meanwhile, on small scales ($k \geq 10 \: h/$Mpc), 
the power in the super-cluster sample is less than 90\% 
of the random sample. 
We then took all members of the two samples 
and found that a correlation exists between the overdensity 
in a sub-volume and the amplitude of the matter power spectrum 
at different scales. In Fig. \ref{fig:pkdmo_overdensity1}, we 
showed that on large scales ($k = 1 \: h/$Mpc)
sub-volumes with higher overdensity generally 
had a higher amplitude, 
whilst on small scales ($k \leq 50 \: h/$Mpc) higher 
overdensity sub-volumes had a lower 
amplitude. 

%matter power spectrum baryons
\item
Our investigation of the effects of baryonic physics on the matter power 
spectrum in our random sample are broadly consistent with previous 
studies \citep[e.g.,][]{2011MNRAS.415.3649V}. 
On intermediate scales ($k = 1-10 \: h/$Mpc) the HYDRO matter power spectrum 
is suppressed by $\sim$10\% in comparison to that of the DMO, while on 
small scales ($k \geq 30 \: h/$Mpc) it is boosted considerably. In addition, we found that 
the effect of baryonic physics on the matter power spectrum does not change significantly 
between $ z = 0 $ and $ z = 1 $ (Fig. \ref{fig:matterpowerspectrumbaryonsredshift}). 
We did find that there is a significant, although very small, difference 
in the effect of baryons on the matter power spectrum between our super-cluster 
and random sample. In general, the effect of baryons on the matter power spectrum 
on all scales  
is suppressed more when the sub-volume 
from which it is being measured is more overdense 
(Figs. \ref{fig:matterpowerspectrumbaryonsredshift}, \ref{fig:matterpowerspectrumbaryons} 
and \ref{fig:pkdmo_overdensity2}). 

%matter power spectrum environment baryons comparison
\item
We compared the selection effect with the baryonic effect on the 
matter power spectrum in Fig. \ref{fig:matterpowerspectrumenvironment}. 
On large scales, specifically for $k\leq \: $5 $h$/Mpc, the effect of 
selection dominates over the effect of baryons, boosting the power. 
On smaller scales than this, the two effects go back and forth between 
complementing and competing with 
one another. However, the selection effect is dominant, 
particularly on scales relevant for radio weak lensing studies. 

%weak lensing environment
\item
Next we studied the effect on the weak lensing convergence 
power spectrum that results from targeting super-cluster regions 
when generating lightcones through the parent simulation. 
We found that the convergence power spectrum measured 
in our super-cluster sample 
had more power on all scales measured, 
than that of our random sample. 
More specifically, for $ \ell > 10^3 $ the amplitude of the 
convergence power spectra for the super-cluster sample was 
twice that of the random sample (Fig. \ref{fig:lightcone_Cl_l}). 

\item
Finally, we made some forecasts for Super-CLASS like weak lensing surveys 
and optical surveys using the same field of view and higher 
galaxy source densities. We found that targeting a super-cluster region 
as opposed to a random region, with the source density expected for the Super-CLASS 
project, generates a detection significance 
of $2.7^{+1.5}_{-1.2}$. 
This indicates that in the absence of systematic effects 
the Super-CLASS project will likely make 
a cosmic shear detection. 
In addition, the radio optical 
cross power spectra generates a detection significance of $7.6^{+4.6}_{-2.7}$, 
therefore cross correlating with an optical survey 
would guarantee a cosmic shear 
detection provided systematics can be 
adequately corrected for (Fig. \ref{fig:lightcone_signal_test_1gal} and 
Table \ref{tab:detection}). 
\end{itemize}

To conclude, although the effect of baryons on the matter power 
spectrum found are important to take into account in future weak lensing surveys, 
they are not a concerning systematic for the Super-CLASS project. 
On scales relevant for radio surveys like Super-CLASS the selection effect is
most important, i.e. the effect of targeting a super-cluster field, 
which boost the amplitude of the 
matter and shear power spectrum considerably. 
Finally, we showed using simulations that the Super-CLASS project's method 
of targeting a super-cluster region of the sky, should allow for a cosmic shear 
detection to be made assuming systematics can be corrected for. 

\section*{Acknowledgements}

We would like to thank Ian McCarthy for providing the BAHAMAS data 
and code, used to create our hydrodynamic simulations. We also thank 
Adrian Jenkins for providing the initial conditions code used in the MACSIS project. 
STK and DJB acknowledge support from STFC through grant ST/L000768/1.
AP and MLB were supported by an ERC Starting Grant (grant no. 280127).

This work used the DiRAC Data Centric system at Durham University, operated by the Institute for Computational Cosmology on behalf of the STFC DiRAC HPC Facility (www.dirac.ac.uk). This equipment was funded by BIS National E-infrastructure capital grant ST/K00042X/1, STFC capital grants ST/H008519/1 and ST/K00087X/1, STFC DiRAC Operations grant ST/K003267/1 and Durham University. DiRAC is part of the National E-Infrastructure.

%%%%%%%%%%%%%%%%%%%%%%%%%%%%%%%%%%%%%%%%%%%%%%%%%%



\bibliographystyle{mnras}
\bibliography{ThesisBibliography} % if your bibtex file is called example.bib

%% Alternatively you could enter them by hand, like this:
%% This method is tedious and prone to error if you have lots of references
%\begin{thebibliography}{99}
%\end{thebibliography}

%%%%%%%%%%%%%%%%%%%%%%%%%%%%%%%%%%%%%%%%%%%%%%%%%%

%%%%%%%%%%%%%%%%% APPENDICES %%%%%%%%%%%%%%%%%%%%%

\appendix

%%%%%%%%%%%%%%%%%%%%%%%%%%%%%%%%%%%%%%%%%%%%%%%%%%

% Don't change these lines
\bsp	% typesetting comment
\label{lastpage}
\end{document}